# Gallium Arsenide (GaAs) Quantum Photonic Waveguide Circuits

Jianwei Wang[1†], Alberto Santamato[1†], Pisu Jiang[1], Damien Bonneau[1], Erman Engin[1], Joshua W. Silverstone[1], Matthias Lermer[2], Johannes Beetz[2], Martin Kamp[2], Sven Höfling[2,3], Michael G. Tanner[4], Chandra M. Natarajan[5], Robert H. Hadfield[4], Sander N. Dorenbos[6], Val Zwiller[6], Jeremy L. O'Brien[1] and Mark G. Thompson[1*]

1. Centre for Quantum Photonics, H H Wills Physics Laboratory and Department of Electrical and Electronic Engineering, University of Bristol, Merchant Venturers Building, Woodland Road, Bristol BS8 1UB, UK
2. Technische Physik and Wilhelm Conrad Röntgen Research Center for Complex Material Systems, Universität Würzburg, Am Hubland, D-97074 Würzburg, Germany
3. SUPA, School of Physics and Astronomy, University of St Andrews, St Andrews, KY16 9SS, UK
4. School of Engineering, University of Glasgow, Glasgow G12 8QQ, UK
5. E. L. Ginzton Laboratory, Stanford University, Stanford 94305, USA
6. Kavli Institute of Nanoscience, TU Delft, 2628CJ Delft, The Netherlands

[†]These authors contributed equally to this work.
* mark.thompson@bristol.ac.uk

## Abstract

Integrated quantum photonics is a promising approach for future practical and large-scale quantum information processing technologies, with the prospect of on-chip generation, manipulation and measurement of complex quantum states of light. The gallium arsenide (GaAs) material system is a promising technology platform, and has already successfully demonstrated key components including waveguide integrated single-photon sources and integrated single-photon detectors. However, quantum circuits capable of manipulating quantum states of light have so far not been investigated in this material system. Here, we report GaAs photonic circuits for the manipulation of single-photon and two-photon states. Two-photon quantum interference with a visibility of 94.9±1.3% was observed in GaAs directional couplers. Classical and quantum interference fringes with visibilities of 98.6±1.3% and 84.4±1.5% respectively were demonstrated in Mach-Zehnder interferometers exploiting the electro-optic Pockels effect. This work paves the way for a fully integrated quantum technology platform based on the GaAs material system.

## 1. Introduction

Quantum information science exploits fundamental quantum mechanical properties – superposition and entanglement – with the goal of dramatically enhancing communication security, computational efficiency and measurement precision [1–4]. Photons have been widely considered as an excellent physical implementation of quantum information and communication technologies due to their low decoherence, fast transmission and ease of manipulation [2, 5]. Bulk optical elements including single-photon sources, single-photon detectors and linear optical circuits have been successfully utilised to experimentally demonstrated quantum communication protocols, quantum metrology and small-scale quantum computation [6–9]. However, this bulk optics approach has severe limitations in terms of circuit stability, complexity and scalability.

The emergence of integrated quantum photonics (IQP) is revolutionising the field of photonic quantum technologies [10]. Utilizing well-developed integration technologies of classical photonics, IQP can shrink quantum experiments from a room-sized optical table onto a coin-sized semiconductor chip, and therefore greatly reduce the footprint of quantum devices and increase the complexity of quantum circuits [5, 11–21]. IQP inherently offers near-perfect mode overlap at an integrated beam splitter for high-fidelity quantum interference [15] and sub-wavelength stability of optical path lengths for high-visibility classical interference [11,14], which are both essential to photonic quantum information processing. Recently, two-photon quantum interference with a visibility of >99%, controlled-NOT quantum gate with a fidelity of 96%, and manipulations of entanglement have been demonstrated in the integrated photonic circuits, based on various platforms such as silica-on-silicon

[11–15], laser direct writing silica [17], lithium niobate [18, 19] and silicon-on-insulator [20, 21], etc. Moreover, IQP would enable on-chip generation, manipulation and detection of quantum states of photons, ultimately required by practical and scalable quantum information processing technologies. Recently, progress also has been made towards integrated single-photon sources and waveguide single-photon detectors. Periodically poled lithium niobate (PPLN) waveguides and silicon wire waveguides as examples of integrated waveguide sources for the generation of photon pairs via spontaneous parametric down conversion (SPDC) and spontaneous four-wave mixing (SFWM) respectively [22, 23]. High-efficiency waveguides superconducting nanowire single-photon detectors (SNSPD) also have been successfully demonstrated in gallium arsenide (GaAs) waveguides and silicon wire waveguides [24, 25].

Here, we report a low-loss GaAs/Al$_{0.3}$Ga$_{0.7}$As ridge waveguide platform for the manipulation of quantum states of light. GaAs is one of the most mature semiconductor materials widely used in classical integrated photonics. GaAs devices have been used for 100 GHz low-power modulation of optical signals [26] based on the strong electro-optical Pockels effect (driven by the large $\chi^2$ nonlinear coefficient of the GaAs material) whose refractive index is linearly proportional to the applied electric field [27], and could provide a route to fast control and manipulation of photons for applications in quantum communication and quantum computation. Moreover, efficient on-chip single-photon sources have been developed based on semiconductor quantum dot embedded in the GaAs photonic crystal waveguides/cavities [28–34]. Spontaneous pair generation techniques have also been investigated using GaAs Bragg-reflection waveguides to achieve the required phase matching condition for spontaneous parametric down conversion [35]. GaAs waveguide integrated superconducting detectors have been demonstrated with efficiencies of 20% [24], short dead time of few ns and photon number resolving capabilities [36]. Recently, photoluminescence from quantum dots has been coupled into the GaAs ridge waveguides and detected using the waveguide SNSPDs [37]. However, to-date no operations of photon's quantum states have been reported in the GaAs waveguide photonic circuits. Based on our GaAs waveguide platform, we demonstrate the ability to control and manipulate two-photon quantum states, demonstrating two-photon quantum interference in directional couplers and utilisng Mach-Zehnder interferometer (MZIs) controlled electro-optically using the Pockels effect to realise quantum interference fringes. This work demonstrates important functionalities required for a GaAs integrated quantum technology platform, and presents essential quantum components for controlling quantum states, opening the way to the monolithic integration of quantum dot/SPDC single-photon sources, quantum photonic circuits and waveguide SNSPDs on a single GaAs device.

2.  **GaAs waveguides and experimental setup**

Fig.1 (a) shows the cross section of a GaAs/Al$_{0.3}$Ga$_{0.7}$As ridge waveguide with a GaAs core and Al$_{0.3}$Ga$_{0.7}$As bottom/top claddings. The refractive indices of the GaAs core and Al$_{0.3}$Ga$_{0.7}$As claddings are 3.431 and 3.282 respectively, at the wavelength of 1550 nm. In order to meet the single-mode condition, the GaAs layer is etched down by 1.5 µm, forming the ridge waveguide with a width of 3.5 µm and a height of 3.9 µm. Fig.1 (a) also shows the simulated field distribution of the transverse electric (TE) fundamental mode using a finite difference mode (FDM) solver. Optical intensity distribution within the fabricated GaAs waveguide has been captured using an infrared CCD camera (Fig.1 (b)), which shows the single mode distribution.

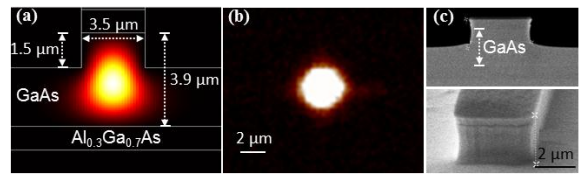

Fig.1. (a) Cross section of the GaAs/Al$_{0.3}$Ga$_{0.7}$As ridge waveguide and its simulated field distribution of the TE fundamental mode at 1550 nm wavelength, (b) measured intensity distribution of the TE fundamental mode at 1550 nm wavelength and (c) Scanning Electron Microscopy (SEM) images of the GaAs/Al$_{0.3}$Ga$_{0.7}$As waveguide.

The Al$_{0.3}$Ga$_{0.7}$As/GaAs/Al$_{0.3}$Ga$_{0.7}$As layers which form the vertical waveguiding structure were alternately grown on top of a (100) GaAs wafer using molecular beam epitaxy. Note that a 100 nm-thin GaAs cap was also grown upon the top cladding to protect the $Al_{0.3}Ga_{0.7}As$ layer against oxidation, and the GaAs substrate under the bottom cladding was doped to reduce the contact resistance. The waveguide circuits were defined by photolithography, using a 50 nm nickel film hard mask and lift-off process. The GaAs layer was inductively coupled plasma (ICP) etched, and the remaining nickel was removed before the chip was planarized by refilling the etched area with lift-off resist. A 200 nm gold film was sputtered after a second photolithography step, and gold contacts were patterned on top of MZI's arms by the lift-off process. Finally, the chip was cleaved for optical fiber coupling and mounted onto a chip holder for electrical connection. Fig.1 (c) shows the Scanning Electron Microscopy (SEM) images of the GaAs waveguides. Directional couplers and MZIs were both fabricated in this waveguide platform. The measured nominal propagation loss and coupling loss (between waveguides and lensed-fibers with a 2.5±0.5 μm spot-diameter) using the Fabry-Perot method [38] was 1.6 dB/cm and 1.5 dB/facet respectively.

Photon pairs at 1550 nm wavelength were generated via type-II SPDC in a periodically poled potassium titanyl phosphate (PPKTP) nonlinear crystal, pumped with a 50 mW continuous-wave laser at 775 nm wavelength (Fig.2). Dichroic mirrors and a long-pass filter were used to separate the bright pump light from the photon pairs. Photon pairs with orthogonal polarization were separated by a polarization beam splitter (PBS) and collected into two polarization-maintaining fibers (PMFs). Photons with horizontal polarization (corresponding to the TE mode of the waveguides) were coupled to the GaAs devices via two lensed single-mode fibers (lensed-SMFs), where the polarization orientation was corrected using two fiber polarization controllers for injection into the test devices. After the chip, photons were collected by two lensed-SMFs and detected using two single-photon detectors. Coincidences were recorded using a Picoharp 300 Time Interval Analyser (TIA). We used two different types of 1550 nm single-photon detectors: 1) two fiber-coupled superconducting single-photon detectors mounted in a closed cycle refrigerator with 1% and 4% efficiencies and ~1kHz dark counts [39], used for the quantum interference experiment in the GaAs directional couplers; 2) two InGaAs/InP Avalanche photodiodes (APDs) from ID Quantique, one working in the free-running mode with a 10% efficiency and the other being gated with a 20% efficiency, for investigation of single-photon superposition state and two-photon entanglement states in the GaAs MZIs. For the APDs, efficiencies and dead time were optimized to balance the coincidence counts and dark counts. A rate of $2\times10^6$ Hz photon pairs from the SPDC source was observed and was used in the following experiments.

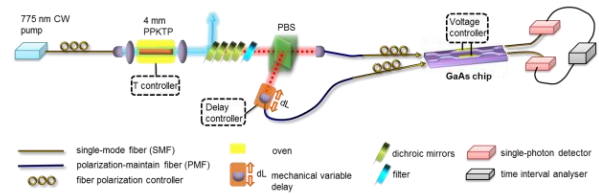

Fig.2. Experimental setup: photon pairs were generated via type-II SPDC in a PPKTP crystal. Photons were collected into two PMFs and coupled to the GaAs chip through two lensed-SMFs, and subsequently routed to two single-photon detectors via another two lensed-SMFs after the chip. Before the chip, a time delay between the two photons was precisely controlled using a mechanical variable delay. A voltage generator was used for electro-optically controlling the relative phase and amplitude of the on-chip photon states.

## 3. Quantum interference

Quantum information encoded on a photon can be realised using any of the different degrees of freedom of a photon, such as path, time, polarization and orbital angular momentum [2, 5]. In path encoding, the qubit is represented using the dual-rail encoding, where a photon in one of the two paths would be defined as $|10\rangle$, and a photon in the other path would be defined as $|01\rangle$. A single-qubit can therefore be represented as a superposition of these two states:

$$|\Psi\rangle = \alpha|10\rangle + \beta|01\rangle \quad (1)$$

where the photon is simultaneously present at $|10\rangle$ and $|01\rangle$ paths with respective probabilities of detection being $|\alpha|^2$ and $|\beta|^2$. The directional coupler (see Fig.3 (a)) is a typical form of integrated beamsplitter and performs a unitary operation of the single-qubit state [11]. Starting with an initial state of $|10\rangle$ for instance, directional coupler rotates it into a superposition state of $\sqrt{1-\varepsilon}|10\rangle + i\sqrt{\varepsilon}|01\rangle$, where $\varepsilon$ is the reflectivity or coupling ratio of the coupler (see details in Appendix A). When the coupling ratio $\varepsilon$ is equal to 0.5, the directional coupler performs a Hadamard-like operation and produces the state $(|10\rangle + i|01\rangle)/\sqrt{2}$. More interestingly, unique quantum interference occurs when two indistinguishable photons meet at a coupler with an $\varepsilon$ of 0.5 [40]. According to the interpretation of quantum mechanics, when two processes are indistinguishable, the probability of an event is equal to the complex square of their added probability amplitudes. Due to the $\pi/2$ phase shift for any photon reflected from a beamsplitter, the probabilities of both photons being reflected or both transmitted cancel out; and therefore two photons injected on a coupler bunch together and produce a maximally path-entangled state as:

$$(|20\rangle + |02\rangle)i/\sqrt{2} \quad (2)$$

When two optical waveguides are placed closely together, light will couple back-and-forth between them via the evanescent field [41]. The coupling ratio $\varepsilon$ of the directional coupler depends on its coupling length and coupling strength. We designed and fabricated GaAs directional couplers with different coupling lengths and gaps for a control of the coupling ratio $\varepsilon$ (Fig.3 (a)). A directional coupler with near 0.5 coupling ratio was obtained when the gap was 2.5 μm and the coupling length was around 140 μm (Fig.3 (b)). The total length of the device was about 7 mm including four S-bends with a radius of 10 mm and the input /output waveguides (of separation 250 μm). Two input/output access-waveguides were distanced by 250 μm to allow access of the lensed-fibers for input/output coupling. At the 1550 nm wavelength, the fiber-to-fiber loss of the chip was measured to be ~9 dB, with the internal devices losses (including the propagation loss and bends loss) estimate to be ~3 dB. To characterise the device in the quantum regime, photon pairs from the SPDC source were launched to the GaAs directional coupler. A variable time delay between the two injected photons was precisely controlled using a mechanical variable delay with a step of 20 μm. After the chip, coincidences detection events were measured using two superconducting detectors and a TIA. Fig.3 (c) shows the Hong-Ou-Mandel (HOM) dip with a visibility (($N_{Max}$-$N_{Min}$)/$N_{Max}$) of 94.9±1.3%, after a subtraction of accidental coincidences [40]. At the dip position, quantum interference results in the two photons coherently bunched together (see formula (2)), and therefore minimal coincidences are recorded there. Observation of the high-visibility HOM-dip experimentally confirms two-photon quantum interference within the GaAs directional coupler. The shape of the HOM-dip is determined by the Fourier transform of the spectrum of the two-photons state. Here, the triangular shape arises from the natural SPDC phase-matching sinc$^2$ spectrum, which is narrower than the bandwidth of filters used in the SPDC source. A triangular fit is used for estimating the visibility and shape of the HOM-dip. The shoulder-to-shoulder width of the HOM-dip is 440 μm, indicating that the coherence time of each photon is 0.73 ps and coherent length in the waveguides is 64.1 μm.

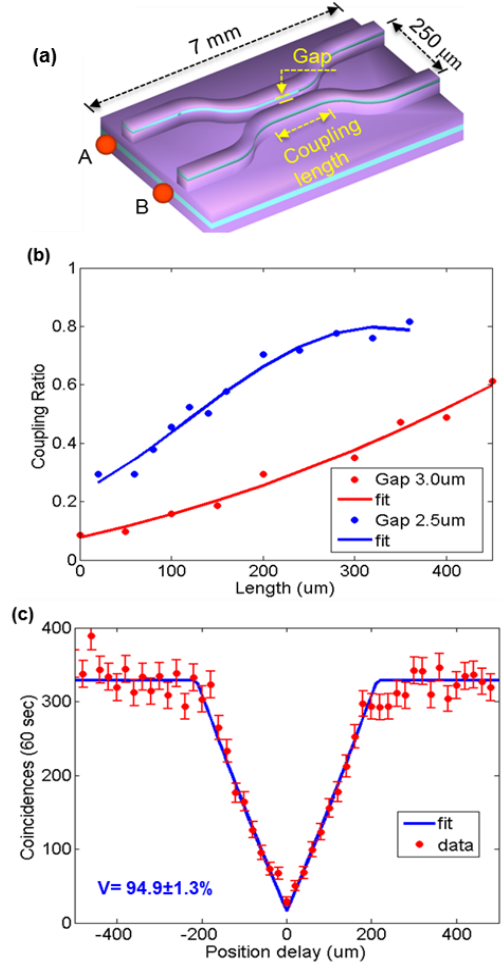

Fig.3. (a) Schematic diagram of the GaAs directional couplers. (b) Measured coupling ratio of the GaAs directional couplers with different gaps as the coupling length increases. Solid lines are fits and points are measured data. (c) Two-photon quantum interference in the GaAs directional coupler with near 0.5 coupling ratio, showing high visibility of 94.9±1.3%. Solid line is an inverse triangular fit for an estimation of the visibility and shape of the HOM-dip. Coincidences were measured using two superconducting detectors with 1% and 4% efficiencies and ~1 Hz dark counts [39]. Accidental coincidences are subtracted and error bars arise from Poissonian statistics.

Furthermore, we measured the indistinguishability of photon pairs directly from the SPDC source using a fiber beam splitter ($\varepsilon = 0.5$) connected with PMFs, resulting in maximum visibility of the HOM-dip of 98.7±0.6%. Compared with the visibility for the GaAs coupler, a 3.8% degradation of the visibility was observed and attributed to the strong Fresnel reflections at the waveguide facets due to the large refractive index difference. Since the coherence length of each photon is much shorter than the distance between the facets, we can ignore the Fabry-Perot self-interference of photons and only consider their back-and-forth reflections between waveguide facets. At each facet between GaAs waveguides and air, photons have an $R$ probability of being reflected and a $T$ probability of being transmitting. $R$ and $T$ are respectively calculated to be 30% and 70% using the Fresnel equations (($n_{GaAs}$-$n_{air}$)/($n_{GaAs}$+$n_{air}$))$^2$, where $n_{GaAs}$ and $n_{air}$ are refractive indices of GaAs and air. Firstly, consider the condition where the two photons undergo quantum interference and bunch at the output ports (i.e. centre of the HOM-dip). Due to the reflections at the output facets photon $A$ transmits with the $T$ probability and photon $B$ is reflected back with the $R$ probability. Photon $B$ can be reflected again at the input facets and leave out from another output port of the coupler with a phase-dependent probability. That is, round-trip reflections result in extra coincidences between photon $A$ and photon $B$ (see Appendix. B), even in the case of perfect quantum interference. Note that time window for coincidences measurement was >5 ns which was much longer than the first-order round-trip time delay of about 80 ps. Then, at the shoulder position of the HOM-dip corresponding to the distinguishable photons pairs input, we can use the same model to estimate the coincidences. Considering the loss within the chip, theoretical degradation of the visibility is estimated to be in the range of 0~4.4%, which depends on the phase difference between two input access-waveguides before the coupler. The experimental 3.8% degradation of visibility is within this theoretical range and actually smaller than the worst degradation owing to a non-zero phase difference. The problem of reflection on facets could be resolved by applying anti-reflection coating on the waveguide facets. In future, for GaAs quantum circuits monolithically integrated with on-chip single-photon sources and detectors, reduction of the visibility due to the facet reflection could be ignored.

## 4. Manipulation of quantum states

Arbitrary unitary operations of quantum states, including preparation, manipulation and measurement of quantum states, are required to implement quantum communication and universal quantum computing. Generally, an arbitrary unitary operator on single-qubit can be decomposed of a set of rotations as $U_{arb}$=exp($i\alpha\sigma_z$/2) exp($i\beta\sigma_y$/2) exp($i\gamma\sigma_z$/2), which physically behaves as one Mach-Zehnder interferometer (MZI) and two additional phase shifters [1]. MZI consisting of two beams splitters and phase shifters is capable of controlling the relative phase and amplitude of the superposition state and entanglement state. When the single-photon state |10⟩ is launched into the MZI, the state is transforms to:

$$[(1 - 2\varepsilon)\cos(\theta/2) + i\sin(\theta/2)] |10\rangle \\ +i2\sqrt{\varepsilon(1-\varepsilon)}\cos(\theta/2) |01\rangle \quad (3)$$

where $\theta$ is the relative phase between two arms and $\varepsilon$ is the coupling ratio of two identical couplers. An MZI with additional phase shifters enables arbitrary operations of the single-qubit states and therefore functionalizes as the building-block for an experimental realization of arbitrary unitary $N\times N$ operators [42] and also for the large-scale quantum information processors [13].

We fabricated GaAs MZIs with two electro-optical phase shifters which enable an independent control of the phases of two arms (Fig.4 (a)). When an electric field $E$ is applied along the (100) direction (vertically), the refractive index of the TE mode linearly responds to the electric field as $\Delta n = n^3_{GaAs} r_{14} E/2$, where $r_{14} \sim 1.4\times 10^{-12}$ m/V is the electro-optical coefficient of the GaAs material [43]. The length of the phase shifters 1.0 cm and voltage required to induce a π phase shift ($V_\pi$) was measured to be 13 V. The two couplers within the MZI were designed identically with a gap of 3.0 μm and a coupling length of 255 μm. The total length of the MZI chip was about 1.7 cm (Fig.4 (b)) and fiber-to-fiber loss of the chip was measured to be -10.3 dB. Classical characterisation of the device was performed using coherent bright-light from a tunable laser diode, and also single-photons from the SPDC source were individually routed to the MZI devices for a characterization of classical interference. Two power-meters and two APDs were respectively used to detect the bright-light intensities and single-photon counts at two output ports of the MZI. By linearly scanning the applied voltages on two arms, we observed the classical interference fringes for both bright-light and single-photons which exhibited the same periodicity. Fig.4 (c) shows the normalized classical interference fringes as a function of relative phase shift for the coherent bright-light input. One can see that the classical interference fringes for two outputs are unbalanced and have different maximum visibilities of 98.6±1.3% and 79.9±4.9%. The unbalance of the interference fringes arises from the non-0.5 coupling ratios of two identical directional couplers. The coupling ratio of individual coupler was measured to be approximately 0.3 (Fig.3 (b)). According to the formula (3), single-photon counts or bright-light intensities from two outputs respectively vary as $\sin^2(\theta/2) + \cos^2(\theta/2)(1-2\varepsilon)$ and $4\varepsilon(1-\varepsilon)\cos^2(\theta/2)$, and we plot the corresponding theoretical fringes when the $\varepsilon$ is chosen to be 0.3 (solid lines in Fig.4 (c)). Theoretical fringes are consistent with the experimental interference fringes. It is anticipated that MZI consisting couplers with near 0.5 coupling ratios would offer sinusoidal outputs as $\sin^2(\theta/2)$ and $\cos^2(\theta/2)$ and therefore result in well-balanced and higher-visibility classical interference fringes. Note that we actually had MZIs with $\varepsilon$ close to 0.5; however, they unfortunately suffered high loss which made it unfeasible to characterise these devices in the two-photon quantum interference experiments. Then we used the MZI with $\varepsilon \sim 0.3$ for investigation of quantum interference within these devices.

Generally, when two indistinguishable photons are separately launched into two input ports of the MZI, quantum interference at the first coupler with an arbitrary coupling ratio creates the two-photon state:

$$\sqrt{2\varepsilon(1-\varepsilon)}\, i(|20\rangle+|02\rangle) + (1-2\varepsilon)\,|11\rangle \qquad (4)$$

(For details see Appendix. A). Note that when $\varepsilon$ is equals to 0.5 the two photons are maximally path-entangled, as in formula (2). The phase shifters within the MXI then perform a $z$-axis rotation on the state, and the second coupler acts to further transform the state to:

$$\sqrt{2\varepsilon(1-\varepsilon)}[-\varepsilon e^{-i2\theta}+(1-2\varepsilon)e^{-i\theta}+1-\varepsilon]i\,|20\rangle +$$
$$\sqrt{2\varepsilon(1-\varepsilon)}[(1-\varepsilon)\,e^{-i2\theta}+(1-2\varepsilon)e^{-i\theta}-\varepsilon]i\,|02\rangle +$$
$$[-2\varepsilon(1-\varepsilon)e^{-i2\theta}+(1-2\varepsilon)^2 e^{-i\theta}-2\varepsilon(1-\varepsilon)]\,|11\rangle$$

To characterise the performance of the device in the two-photon quantum regime, we routed photons pairs from the SPDC source to the MZI and recorded coincidences (corresponding to the $|11\rangle$ term in the formula (5)) using two APDs and the TIA. The time delay between the two photons was carefully controlled to make them arrive at the MZI simultaneously and therefore guarantee the time-indistinguishability. Compared with the classical interference fringes above, two-photon quantum interference fringe with a double frequency was observed and shown in Fig. 4(d), indicating a manipulation of the two-photon entanglement state. The maximum visibility is measured to be 84.4±1.5 %, which is greater than the requirement of beating the standard quantum limit [44]. The visibility of quantum interference fringe is non-uniform owing to the unbalance of the directional couplers [45], and in the classical interference fringes. The coexistence of the $e^{-i\theta}$ and $e^{-i2\theta}$ terms in the formula (5) leads to the non- uniformity of the interference fringe when $\varepsilon$ is away 0.5. The solid line in Fig.4 (d) is the theoretical two- photon interference

$$+ \cos(\theta) |11\rangle$$

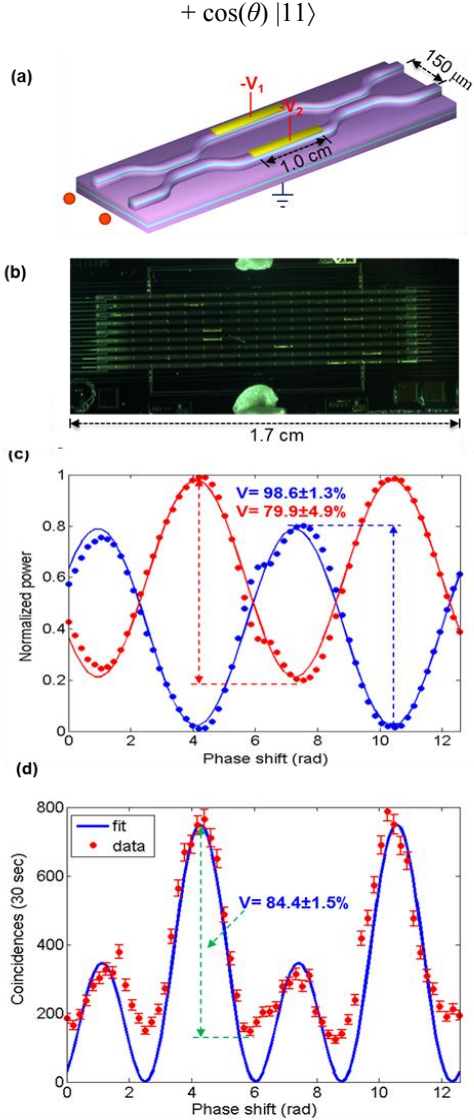

Fig.4. (a) schematic diagram of the GaAs Mach-Zehnder interferometer (MZI) with two directional couplers and two electro-optical phase shifters. (b) Optical microscopy image of the fabricated GaAs MZIs. (c) Classical interference fringes. Normalized intensities of two outputs are plotted as a function of relative phase shift for the coherent bright-light input (the same periodicity as single-photons input). (d) Quantum interference fringe showing a manipulation of the two-photon state. Coincidences are plotted as a function of relative phase shift for the indistinguishable photons pair input. Solid lines in (c) and (d) are theoretical fringes when the $\varepsilon$ of two couplers is 0.30. Coincidences were measured using two APDs and the TIA. Accidental coincidences are subtracted and error bars arise from Poissonian statistics.

fringe when the $\varepsilon$ is chosen as 0.3. The shape and periodicity between the theoretical fringe and experimental result agree well, whereas deviation at the bottom likely arises from the polarization distinguishability induced in SMFs before the chip. Additionally, two photons may a carry small transverse magnetic (TM) component, which does not response to the applied electric field, and behave as the coincidences background independent of the phase shift. If a coupling ratios of 0.5 was used, the formula (5) can be simply reduces to $\sin(\theta)(|20\rangle - |02\rangle)/\sqrt{2} + \cos(\theta) |11\rangle$, resulting in pure double-frequency quantum interference fringe [13, 14, 20-22]. Through further device optimisation, controlling coupling ratios and polarization of photons, quantum interference with uniform distribution and higher visibility would be achievable.

## 5. Conclusion

To summarize, we have developed a GaAs ridge waveguide technology platform for integrated quantum photonic circuits. Directional couplers and MZIs were fabricated and their suitability for quantum interference experiments assessed. We demonstrated two-photon quantum interference with a high visibility using the directional couplers and implemented the manipulation of two-photon state using MZIs. This study demonstrates the feasibility of quantum waveguide circuits in GaAs, opening the way to a fully integrated quantum technology platform where single photon sources, detectors and waveguide circuits could be combined in a single GaAs chip. This approach is promising for a large-scale and practical integrated platform for on-chip quantum information processing.

## Acknowledgements

We thank X-Q. Zhou, P. Shadbolt, J.C.F Matthews and X. Cai for discussions. This work was supported by the European FP7 project Quantum Integrated Photonics (QUANTIP), the Engineering and Physical

Sciences Research Council (EPSRC), the European Research Council (ERC) and the Bristol Centre for Nanoscience and Quantum Information (NSQI). A.S. is supported by a DSTL National PhD scheme studentship. R.H.H. acknowledges a Royal Society University Research Fellowship. V.Z. acknowledges support from the Dutch Foundation for Fundamental Research on Matter. J.L.O'B. acknowledges a Royal Society Wolfson Merit Award.

## Appendix. A

Unitary operator of the directional coupler with an arbitrary coupling ratio or reflectivity $\varepsilon$ is shown as:

$$\begin{bmatrix} \sqrt{1-\varepsilon} & i\sqrt{\varepsilon} \\ i\sqrt{\varepsilon} & \sqrt{1-\varepsilon} \end{bmatrix} \quad \text{A. (1)}$$

We use the quantum mechanical representation to describe the unitary transformations applied by the directional coupler and Mach-Zehnder interferometer (MZI) as following. $\hat{a}_i$ and $\hat{a}_i^+$ are the annihilation and creation operators, respectively, and $i$ is the port number in Fig. A.1.

When one-photon state $|10\rangle = \hat{a}_1^+|00\rangle$ is launched into the directional coupler, the state is rotated as:

$$|10\rangle \xrightarrow{DC} (\sqrt{1-\varepsilon}\,\hat{a}_3^+ + i\sqrt{\varepsilon}\,\hat{a}_4^+)\,|00\rangle$$
$$= \sqrt{1-\varepsilon}\,|10\rangle + i\sqrt{\varepsilon}\,|01\rangle \quad \text{A. (2)}$$

When one-photon state $|10\rangle = \hat{a}_1^+|00\rangle$ is launched into the MZI, the state is rotated as:

$$|10\rangle \xrightarrow{DC1} (\sqrt{1-\varepsilon}\,\hat{a}_3^+ + i\sqrt{\varepsilon}\,\hat{a}_4^+)\,|00\rangle$$
$$\xrightarrow{\text{Phase shifter}} (\sqrt{1-\varepsilon}\,\hat{a}_3^+ + e^{-i\theta}i\sqrt{\varepsilon}\,\hat{a}_4^+)\,|00\rangle$$
$$\xrightarrow{DC2} [\sqrt{1-\varepsilon}\,(\sqrt{1-\varepsilon}\,\hat{a}_5^+ + i\sqrt{\varepsilon}\,\hat{a}_6^+) +$$
$$e^{-i\theta}i\sqrt{\varepsilon}\,(i\sqrt{\varepsilon}\,\hat{a}_5^+ + \sqrt{1-\varepsilon}\,\hat{a}_6^+)]\,|00\rangle$$
$$= [(1-2\varepsilon)\cos(\theta/2) + i\sin(\theta/2)]\,|10\rangle +$$
$$i2\sqrt{\varepsilon(1-\varepsilon)}\cos(\theta/2)\,|01\rangle \quad \text{A. (3)}$$

When two-photon state $|11\rangle = \hat{a}_1^+\hat{a}_2^+|00\rangle$ is launched into the directional coupler, the state is rotated as:

$$|11\rangle \xrightarrow{DC} (\sqrt{1-\varepsilon}\,\hat{a}_3^+ + i\sqrt{\varepsilon}\,\hat{a}_4^+)(i\sqrt{\varepsilon}\,\hat{a}_3^+ + \sqrt{1-\varepsilon}\,\hat{a}_4^+)\,|00\rangle$$
$$= [i\sqrt{\varepsilon(1-\varepsilon)}\,(\hat{a}_3^+\hat{a}_3^+ + \hat{a}_4^+\hat{a}_4^+) + (1-2\varepsilon)\hat{a}_3^+\hat{a}_4^+]\,|00\rangle$$
$$= \sqrt{2\varepsilon(1-\varepsilon)}\,i(|20\rangle + |02\rangle) + (1-2\varepsilon)\,|11\rangle \quad \text{A. (4)}$$

When two-photon state $|11\rangle = \hat{a}_1^+\hat{a}_2^+|00\rangle$ is launched into the MZI, the state is rotated as:

$$|11\rangle \xrightarrow{DC1} [i\sqrt{\varepsilon(1-\varepsilon)}\,(\hat{a}_3^+\hat{a}_3^+ + \hat{a}_4^+\hat{a}_4^+) +$$
$$(1-2\varepsilon)\hat{a}_3^+\hat{a}_4^+]\,|00\rangle$$
$$\xrightarrow{\text{Phase shifter}} [i\sqrt{\varepsilon(1-\varepsilon)}\,(\hat{a}_3^+\hat{a}_3^+ + e^{-i2\theta}\hat{a}_4^+\hat{a}_4^+) +$$
$$(1-2\varepsilon)e^{-i\theta}\hat{a}_3^+\hat{a}_4^+]\,|00\rangle$$
$$\xrightarrow{DC2} i\sqrt{2\varepsilon(1-\varepsilon)}[-\varepsilon e^{-i2\theta} + (1-2\varepsilon)e^{-i\theta} +$$
$$1-\varepsilon]\,|20\rangle + i\sqrt{2\varepsilon(1-\varepsilon)}[(1-\varepsilon)\,e^{-i2\theta} +$$
$$(1-2\varepsilon)e^{-i\theta} - \varepsilon]\,|02\rangle +$$
$$[-2\varepsilon(1-\varepsilon)e^{-i2\theta} + (1-2\varepsilon)^2 e^{-i\theta} - 2\varepsilon(1-\varepsilon)]\,|11\rangle$$
$$\quad \text{A. (5)}$$

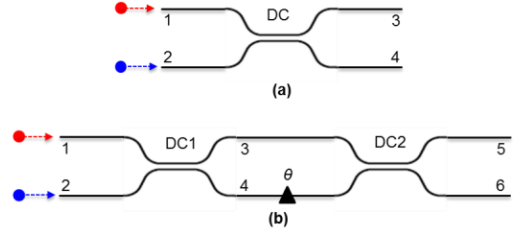

Fig. A.1. Schematic diagrams of the (a) directional coupler and (b) Mach-Zehnder interferometer.

## Appendix. B

Coherence length of each photon is much shorter than the chip length and therefore we could ignore the Fabry-Perot self-interference and only consider the forth-and-back reflections for photons. At each waveguide facet, photons have a $R$ probability of being reflected and a $T$ probability of transmitting. $R$ and $T$ are estimated using the Fresnel equation.

Fig.B.1. (a) shows the forth-and-back reflection at the tip position of the HOM-dip, where two photons should be coherently bunched in the idea case. For example, photon $A$ (red) and photon $B$ (blue) have a 50% probability of being bunched at the port 3. Due to the facet reflection, photon $A$ have the $T$ probability of transmitting and photon $B$ has the $R$ probability of being reflected. Then photon $B$ is reflected again at the input facets, and consequently there is a chance that photon $B$ will leave out from the port 4 of the coupler and coincide with photon $A$. Note that photon $B$ passes the directional coupler ($\varepsilon=0.5$) twice and a "MZI"-like interference will occur. Any variation of waveguides width/length and angled-cleave of the input access-waveguides will induce phase difference $\Delta\varphi$ between two input waveguides before the coupler. Therefore, the probability of extra coincidences depends on the phase difference $\Delta\varphi$. When only considering the first-order round-trip of reflections, coincidences at the dip position will be:

$$NR^2T^4\eta_c^4\eta^4\cos^2(\Delta\varphi/2)\eta_{d1}\eta_{d2} \quad \text{B. (1)}$$

where $N$ is the rate of photon pairs of the SPDC source, and $\eta_c$ is the coupling loss and and $\eta$ is the loss within the chip (including the propagation loss and bending loss), and $\eta_{d1}$ and $\eta_{d2}$ are efficiencies of two detectors.

Similarly, we can analysis the shoulder of the HOM-dip, where two distinguishable photons are injected and four different processes occur: both reflected, both transmitted, and one reflected and one transmitted. Fig.B.1. (b) and (c) show the zero-order and first-order round-trips when photon $A$ and $B$ are initially antibunched. Coincidences at the shoulder position of the HOM-dip will be:

$$N [T^4\eta_c^4\eta^2/2 + \sin^2(\Delta\varphi/2)R^2T^4\eta_c^4\eta^4/2 + \cos^2(\Delta\varphi/2)R^2T^4\eta_c^4\eta^4/2] \eta_{d1}\eta_{d2} \qquad \text{B. (2)}$$

According to the formulas B. (1) and (2), the theoretical visibility is estimated to be in the range of 95.6%~100%, corresponding to a degradation of the visibility in the range of 0~4.4%, which depends on the phase difference $\Delta\varphi$ between two input access-waveguides before the coupler. The worst degradation of the visibility is 4.4% when the $\Delta\varphi$ is chosen to be zero.

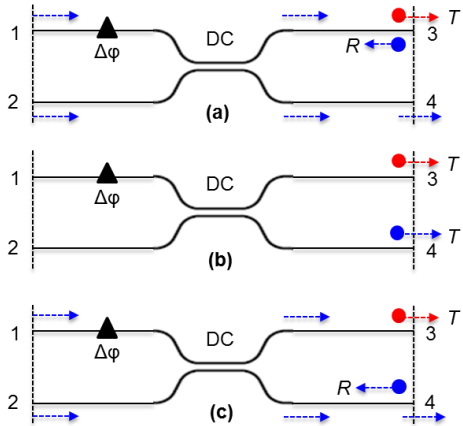

Fig. B.1. Illustration of round-trip reflections of photons in the directional coupler ($\varepsilon=0.5$).